\documentclass{webofc}

\usepackage[varg]{txfonts}   
\usepackage{graphicx}
\usepackage{amsmath}
\usepackage{amssymb}
\usepackage{tikz}
\usetikzlibrary{positioning,fit,backgrounds,shapes.geometric,arrows.meta,shadows.blur,calc,decorations.pathreplacing}
\usepackage{hyperref}
\usepackage{url}
\usepackage{ragged2e}   
\RequirePackage{orcidlink}
\hypersetup{colorlinks=true,citecolor=blue,urlcolor=blue,linkcolor=blue}

\newcommand{\pt}{p_{\mathrm{T}}}
\newcommand{\ttbar}{t\bar{t}}
\newcommand{\avgmu}{\langle\mu\rangle}

\begin{document}

\title{Using Line Segment Tracking to enable ML-based Track Reconstruction in CMS}

\author{\firstname{Aashay} \lastname{Arora}\inst{1}\orcidlink{0000-0003-3453-4740}\thanks{\email{aashay.arora@cern.ch}}
        on behalf of the CMS Collaboration\thanks{Copyright 2026 CERN for the benefit of the CMS Collaboration. Reproduction of this article or parts of it is allowed as specified in the CC-BY-4.0 license.}
}

\institute{University of California San Diego, La Jolla, CA, USA}

\abstract{%
High-pileup conditions in CMS during the HL-LHC era make charged-particle tracking
increasingly challenging as detector occupancy and combinatorics grow. We present a
hybrid approach that exploits Line Segment Tracking (LST) objects rather than
individual hits to enable the first CMS ML-based track reconstruction algorithm. The
LST objects, namely the triplets used in this study, are built according to
geometry- and physics-driven criteria and carry
richer local structure, geometrical cues, and physics quantities such as transverse
momentum, which helps the model incorporate detector-level information from the start.
We apply a Transformer encoder with geometrically masked sparse self-attention to embed
these objects in a learned latent space, and use object condensation to assemble full
tracks in a single inference step. This object-based representation reduces complexity
in dense environments and makes the model less sensitive to ambiguities introduced by
high pileup. Evaluated on CMS Phase-2 simulation of $\ttbar$ events at $\avgmu = 200$,
the approach achieves reconstruction efficiency comparable to the LST quintuplet
baseline while reducing the fake rate by roughly a factor of three and suppressing
duplicates almost entirely, demonstrating the promise of advanced ML architectures for
next-generation particle tracking.%
}
\maketitle

\section{Introduction}
\label{sec:intro}

The High-Luminosity LHC (HL-LHC), at its limit, will deliver an average of about $200$ simultaneous
proton--proton interactions per bunch crossing. Under these conditions charged-particle
tracking becomes the single most expensive step of event reconstruction in CMS: Kalman filter based algorithms
that underpin the current track-building sequence are
intrinsically sequential, and its cost grows steeply with hit density. Meeting the
latency budget of the Phase-2 High-Level Trigger (HLT) therefore requires pattern
recognition that is both cheaper and naturally parallel.

Line Segment Tracking (LST)~\cite{lst_arxiv,CMS-DP-2025-051} addresses this by
restructuring track finding in the Phase-2 Outer Tracker (OT) into a hierarchy of purely
local construction steps. It is therefore intrinsically parallelizable and
hardware-agnostic: objects of the same type are mutually independent and can be built
simultaneously across the entire detector, and the C++ implementation, using the Alpaka
portability layer~\cite{alpaka}, runs unmodified on both CPU and GPU. LST is integrated
into CMSSW and is established for Phase-2 HLT tracking~\cite{CMS-DP-2025-051}, with
machine learning already applied within it for object selection and duplicate
removal~\cite{CMS-DP-2023-075,CMS-DP-2025-048}.

In parallel, machine learning (ML) has become highly successful across the event
reconstruction chain: the Exa.TrkX~\cite{exatrkx2021} pipeline, for instance, has shown
that learned models can perform track finding at HL-LHC occupancies. A recurring
difficulty is that most such models build tracks directly from individual detector hits,
without any prior physics information, and must therefore learn the effects of the
magnetic field and of the detector geometry from scratch. An attractive alternative is to
\emph{tokenize} the hits into larger, physically meaningful objects. The intermediate
objects that LST already constructs, in particular its triplets, are precisely such a
tokenization: each carries a position, a direction and a local transverse-momentum
estimate derived from the correlated hit pairs of the Phase-2 $\pt$ modules.

This contribution presents the first CMS demonstration of a track-reconstruction model
built on this idea~\cite{CMS-DP-2026-030}. LST triplets are used as input tokens to a
Transformer encoder with geometrically masked sparse self-attention, trained under an
object condensation objective~\cite{kieseler2020objectcondensation,lieret2023highpileup};
the resulting embeddings are clustered with DBSCAN~\cite{ester1996dbscan} to form track
candidates in a single inference step.

\section{Line Segment Tracking objects as tokens}
\label{sec:lst}

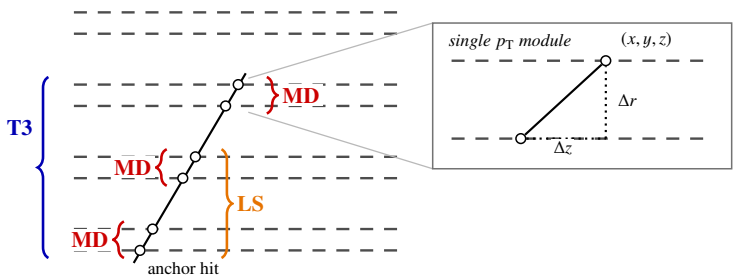
\begin{figure}[htbp]
    \centering
    \resizebox{0.74\textwidth}{!}{%
    \begin{tikzpicture}[
        font=\small, scale=1.02, every node/.style={transform shape},
        lay/.style={dash pattern=on 6pt off 5pt, line width=1.1pt, black!70},
        hit/.style={circle, draw=black, fill=white, line width=0.7pt, inner sep=0pt, minimum size=4.4pt},
        seg/.style={line width=1.0pt},
        br/.style={decorate, decoration={brace, amplitude=5pt, raise=1pt}, line width=1.2pt},
        lead/.style={gray!55, line width=0.6pt}
    ]
    \foreach \y in {0, 1.15, 2.30, 3.45} {
        \draw[lay] (-0.15,\y)      -- (5.10,\y);
        \draw[lay] (-0.15,\y+0.34) -- (5.10,\y+0.34);
    }
    \draw[seg] (0.80,-0.20) -- (2.58,2.82);
    \node[hit] (h1a) at (0.90,0.00) {};
    \node[hit] (h1b) at (1.10,0.34) {};
    \node[hit] (h2a) at (1.58,1.15) {};
    \node[hit] (h2b) at (1.78,1.49) {};
    \node[hit] (h3a) at (2.26,2.30) {};
    \node[hit] (h3b) at (2.46,2.64) {};
    \node[font=\footnotesize, anchor=north west, fill=white, inner sep=1pt] at (0.98,-0.14) {anchor hit};
    \draw[br, red!80!black] (0.68,-0.12) -- (0.68,0.46) node[midway, left=6pt, font=\bfseries, fill=white, inner sep=1.5pt] {MD};
    \draw[br, red!80!black] (1.36,1.03) -- (1.36,1.61) node[midway, left=6pt, font=\bfseries, fill=white, inner sep=1.5pt] {MD};
    \draw[br, red!80!black] (2.88,2.76) -- (2.88,2.18) node[midway, right=6pt, font=\bfseries, fill=white, inner sep=1.5pt] {MD};
    \draw[br, orange!90!black] (2.16,1.61) -- (2.16,-0.12) node[midway, right=6pt, font=\bfseries, fill=white, inner sep=1.5pt] {LS};
    \draw[br, blue!70!black] (-0.52,-0.12) -- (-0.52,2.76) node[pos=0.72, left=6pt, font=\bfseries, fill=white, inner sep=1.5pt] {T3};
    \draw[lead] (2.62,2.74) -- (5.55,3.62);
    \draw[lead] (2.62,2.20) -- (5.55,1.30);
    \draw[black!55, line width=0.7pt] (5.55,1.30) rectangle (10.30,3.62);
    \draw[lay] (5.85,1.78) -- (10.00,1.78);
    \draw[lay] (5.85,3.02) -- (10.00,3.02);
    \node[hit] (za) at (6.95,1.78) {};
    \node[hit] (zb) at (8.30,3.02) {};
    \draw[seg] (za) -- (zb);
    \draw[dotted, line width=1.0pt] (za) -- (8.30,1.78);
    \draw[dotted, line width=1.0pt] (8.30,1.78) -- (zb);
    \node[below=0pt, font=\footnotesize, fill=white, inner sep=1pt] at (7.62,1.78) {$\Delta z$};
    \node[right=2pt, font=\footnotesize] at (8.30,2.40) {$\Delta r$};
    \node[font=\footnotesize, anchor=south west] at (8.42,3.10) {$(x,y,z)$};
    \node[font=\footnotesize\itshape, anchor=north west] at (5.68,3.56) {single $\pt$ module};
    \end{tikzpicture}%
    }
    \caption{Construction of an LST T3 and its underlying objects in the Phase-2 Outer
    Tracker, shown in the $r$--$z$ plane}
    \label{fig:lstobjects}
\end{figure}

LST is designed around a distinctive feature of the Phase-2 OT geometry. Each layer is
instrumented with $\pt$ modules of two closely spaced silicon sensors, so a charged
particle deposits a hit in each; because the trajectory is bent in the $3.8$~T solenoidal
field, the displacement between the two hits encodes the local bending and hence a coarse
estimate of its transverse momentum. Every OT layer therefore supplies not just a
position but a directional and momentum constraint, which LST exploits at every level of
a bottom-up hierarchy of objects, each formed from those of the preceding level using
only \emph{local} geometric information (Figure~\ref{fig:lstobjects}).
\textbf{Mini-doublets} (MD) link the two hits of a single $\pt$ module when their
displacement is consistent with a track above $0.8$~GeV, the offline counterpart of the
on-detector stub. \textbf{Line segments} (LS) join two MDs in neighbouring layers whose
positions and directions are compatible with a single trajectory, with the allowed
pairings fixed in advance by precomputed module connection maps. \textbf{Triplets} (T3)
merge two LSs sharing a middle MD into a three-layer object whose curvature gives a
refined local $\pt$ estimate, and \textbf{quintuplets} (T5) combine two T3s sharing a
middle MD into a five-layer object, the principal pure-OT track candidate.

Inner Tracker pixel seeds, reconstructed in the spirit of the Patatrack
workflow~\cite{patatrack}, additionally enter as pixel line segments and are matched to
T3s and T5s to form the pixel-matched objects that complete the LST track-candidate list.
Triplets serve as the tokens for the model described below: they are already rich in
physics content while remaining combinatorially manageable enough to preserve the
flexibility to form longer candidates. Quintuplets, as built by LST
before the track-candidate step, are the reference for the learned candidates.

\section{Model and training}
\label{sec:model}

\subsection{Input representation}
\label{sec:inputs}

Each event is an unordered point cloud of T3 objects. Each mini-doublet is represented by
its \emph{anchor hit}, the hit in the better-resolved sensor of the module: the macro-pixel
P-side in PS modules, whereas in 2S modules both sensors are strips and the resolution is
the coarsest in the OT~\cite{cms_tracker_tdr}. Every triplet is then described by $49$
features, the quantities indicated in Figure~\ref{fig:lstobjects}: for each of its three
mini-doublets, the $(x,y,z)$ coordinates of both hits, the $\eta$ and $\phi$ of the anchor
hit, and the intra-mini-doublet differences $(\Delta\phi, \Delta\phi', \Delta z)$; for each
of its two line segments, $\Delta\phi$, $\Delta\phi'$ and the angles
$(\Delta\alpha_{\mathrm{inner}}, \Delta\alpha_{\mathrm{outer}},
\Delta\alpha_{\mathrm{inner\mbox{-}outer}})$ between the segment direction and that of its
inner mini-doublet, of its outer mini-doublet, and between the two mini-doublet directions;
and the $\pt$, $\eta$ and $\phi$ of the triplet itself, where $\eta$ is taken from the
outermost anchor hit and $\phi$ from the innermost.

In a $\avgmu = 200$ event approximately $10^{4}$ T3s are constructed, of which roughly
$50\%$ are fake, defined as triplets in which not all three mini-doublets originate from the same simulated particle. 
This large fake fraction is not an obstacle but a handle: the object condensation objective assigns such triplets a low
condensation likelihood, providing a natural first-pass fake-rejection step before
clustering.

\subsection{Transformer encoder with geometric sparse attention}
\label{sec:arch}

The Transformer is well matched to this problem: it processes variable-size,
permutation-invariant sets and maps efficiently onto modern GPUs. Its self-attention
mechanism~\cite{vaswani2017attention} maps tokens to queries, keys and values and updates
each token as a learned, data-dependent average over the others,
$\mathrm{Attention}(Q, K, V) = \mathrm{softmax}(Q K^{\mathsf{T}}/\sqrt{d_k})\,V$.

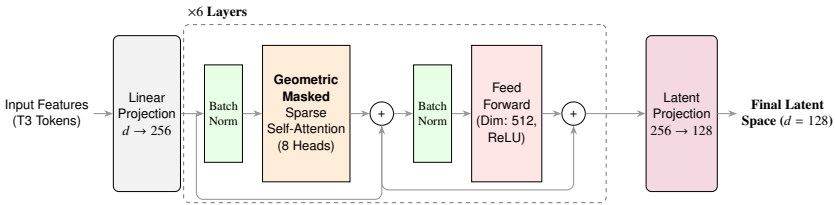
\begin{figure}[htbp]
    \centering
    \resizebox{0.85\textwidth}{!}{%
    \begin{tikzpicture}[
        font=\sffamily,
        >=Stealth,
        node distance=0.8cm,
        layer/.style={draw, rounded corners, fill=blue!10, minimum width=1cm, minimum height=4cm, align=center},
        norm/.style={draw, fill=green!10, minimum width=0.6cm, minimum height=2.5cm, font=\small, align=center},
        attn/.style={draw, fill=orange!15, minimum width=1.2cm, minimum height=3.5cm, align=center},
        ffn/.style={draw, fill=red!10, minimum width=1.2cm, minimum height=3.5cm, align=center},
        connection/.style={->, thick, gray!80},
        add/.style={circle, draw, fill=white, inner sep=0pt, minimum size=6mm, node contents={+}}]
        \node (input) [draw=none, align=center] {Input Features\\(T3 Tokens)};
        \node (lin_proj) [layer, right=0.5cm of input, fill=gray!10] {Linear\\Projection\\$d \to 256$};
        \draw[connection] (input) -- (lin_proj);
        \node (start_block) [coordinate, right=0.4cm of lin_proj] {};
        \node (norm1) [norm, right=0.2cm of start_block] {Batch\\Norm};
        \node (mha) [attn, right=0.5cm of norm1] {\textbf{Geometric}\\\textbf{Masked}\\Sparse\\Self-Attention\\(8 Heads)};
        \node (add1) [add, right=0.5cm of mha];
        \draw[connection] (lin_proj) -- (start_block) -- (norm1);
        \draw[connection] (norm1) -- (mha);
        \draw[connection] (mha) -- (add1);
        \draw[thick, gray!80, ->, rounded corners=5pt] (start_block) -- ++(0,-2.2) -| (add1);
        \node (norm2) [norm, right=0.5cm of add1] {Batch\\Norm};
        \node (ffn) [ffn, right=0.5cm of norm2] {Feed\\Forward\\(Dim: 512,\\ReLU)};
        \node (add2) [add, right=0.5cm of ffn];
        \draw[connection] (add1) -- (norm2);
        \draw[connection] (norm2) -- (ffn);
        \draw[connection] (ffn) -- (add2);
        \draw[thick, gray!80, ->, rounded corners=5pt] (add1.south) -- ++(0,-1.65) -| (add2);
        \begin{scope}[on background layer]
            \node (block_fit) [draw, dashed, thick, gray, rounded corners, fit=(norm1) (add2) (mha), inner sep=15pt] {};
        \end{scope}
        \node [anchor=south west, font=\bfseries, text=black] at (block_fit.north west) {$\times 6$ Layers};
        \node (final_proj) [layer, fill=purple!15, right=1.0cm of block_fit] {Latent\\Projection\\$256 \to 128$};
        \draw[connection] (add2) -- (final_proj);
        \node (output) [right=0.5cm of final_proj, font=\bfseries, align=center] {Final Latent\\Space ($d=128$)};
        \draw[connection] (final_proj) -- (output);
    \end{tikzpicture}%
    }
    \caption{Architecture of the Transformer encoder with geometric sparse attention.}
    \label{fig:model}
\end{figure}

The model, shown in Figure~\ref{fig:model}, projects the input features to a hidden
dimension of $256$ and maps them through six self-attention blocks, each with eight
attention heads and a $512$-dimensional feed-forward sublayer, down to a final
$128$-dimensional latent space. These values were selected using a systematic hyperparameter
scan, which identified the architectural scale yielding the lowest values across all
components of the multi-objective loss.

This operation scales quadratically with the number of tokens, which is prohibitive at
the scale of $10^{4}$ T3s per event. Attention is therefore sparsified by a
geometrically motivated mask: a given T3 attends only to other T3s whose closest hits lie
within $\Delta R < 0.1$ in $\eta$--$\phi$ space. Concretely, for each token $i$ a
neighbour set is defined as
\begin{equation}
    \mathcal{N}(i) = \left\{\, j : \Delta R_{ij} \le \Delta R_{\max}\,\right\},
    \qquad
    \Delta R_{ij} = \sqrt{\left(\eta_i - \eta_j\right)^2 + \Delta\phi_{ij}^2},
    \label{eq:neighbors}
\end{equation}
To bound the memory footprint the set is truncated to the $k$ nearest neighbours in $\Delta R$; a token with no neighbour inside
$\Delta R_{\max}$ attends only to itself, so $\mathcal{N}(i)$ is never empty.

Attention is then evaluated only over these sets. For head $h$, with per-token query, key
and value vectors $\mathbf{q}_i^{(h)}, \mathbf{k}_i^{(h)}, \mathbf{v}_i^{(h)} \in
\mathbb{R}^{d_h}$ of dimension $d_h = d/H$, the softmax normalization runs over
$\mathcal{N}(i)$ rather than over all tokens,
\begin{equation}
    \alpha_{ij}^{(h)} = \frac{\exp\!\left(\mathbf{q}_i^{(h)}\cdot\mathbf{k}_j^{(h)}/\sqrt{d_h}\right)}
    {\displaystyle\sum_{j'\in\mathcal{N}(i)} \exp\!\left(\mathbf{q}_i^{(h)}\cdot\mathbf{k}_{j'}^{(h)}/\sqrt{d_h}\right)},
    \qquad j \in \mathcal{N}(i),
    \label{eq:sparse_softmax}
\end{equation}
with the per-head outputs $\sum_{j\in\mathcal{N}(i)}\alpha_{ij}^{(h)}\mathbf{v}_j^{(h)}$
concatenated over heads and passed through a learned output projection as usual. Because
each token attends to at most $k \ll N$ others, the cost of Eq.~\eqref{eq:sparse_softmax}
scales as $\mathcal{O}(Nk)$ rather than $\mathcal{O}(N^2)$, recovering full attention in
the limit $\Delta R_{\max}\to\infty$. The locality constraint reflects the physical
expectation that triplets belonging to the same track are geometrically proximate, so the
reduction sacrifices no physically relevant context.

\subsection{Object condensation}
\label{sec:oc}

The number of tracks in an event is not known in advance and varies event by event, so
the network output must be converted into a variable-length list of objects. Object
condensation~\cite{kieseler2020objectcondensation,lieret2023highpileup}
is a loss formulation designed for exactly this setting, and is adopted here unchanged.
Each input point is embedded in a low-dimensional latent space and additionally assigned
a \emph{condensation charge} $\beta_i \in [0,1]$, the likelihood that it acts as the
representative, or condensation point, of an object. The objective or loss function, defined explicitly in 
Eq.~(11) of Ref.~\cite{kieseler2020objectcondensation},
combines an attractive term pulling the constituents of an object towards its condensation point, a repulsive
term separating points of different objects, and a term that rewards a high charge on
each condensation point while suppressing the charge of \emph{noise} points, those
belonging to no true object, such as fake triplets.

Minimizing this loss organizes a chaotic cloud of triplets into well-separated clusters,
each anchored by a single condensation point per particle by construction, while learning
a per-token charge that serves as a built-in fake filter.

\subsection{Training and inference}
\label{sec:training}

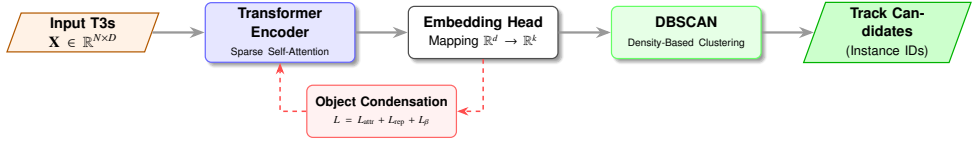
\begin{figure}[htbp]
    \centering
    \resizebox{0.98\textwidth}{!}{%
    \begin{tikzpicture}[
        node distance=1.0cm and 1.2cm,
        process/.style={rectangle, draw=blue!60, fill=blue!10, text width=3.2cm, align=center, minimum height=1.2cm, rounded corners, thick, font=\sffamily, blur shadow={shadow blur steps=5}},
        layer/.style={rectangle, draw=black!70, fill=white, text width=3.2cm, align=center, minimum height=1.2cm, rounded corners, thick, font=\sffamily, blur shadow={shadow blur steps=5}},
        data/.style={trapezium, trapezium left angle=70, trapezium right angle=110, draw=orange!70!black, fill=orange!10, text width=2.6cm, align=center, thick, font=\sffamily},
        loss_style/.style={rectangle, draw=red!60, fill=red!5, text width=3.2cm, align=center, minimum height=1.2cm, rounded corners, thick, font=\sffamily\small}
    ]
    \node (input) [data] {\textbf{Input T3s} \\ $\mathbf{X} \in \mathbb{R}^{N \times D}$};
    \node (trans) [process, right=of input] {\textbf{Transformer Encoder} \\ \scriptsize Sparse Self-Attention};
    \node (embed) [layer, right=of trans] {\textbf{Embedding Head} \\ \small Mapping $\mathbb{R}^d \to \mathbb{R}^k$};
    \node (dbscan) [process, right=of embed, fill=green!10, draw=green!60] {\textbf{DBSCAN} \\ \scriptsize Density-Based Clustering};
    \node (output) [data, right=of dbscan, fill=green!20, draw=green!70!black] {\textbf{Track Candidates} \\ \small (Instance IDs)};
    \node (loss) [loss_style, below=1.2cm of $(trans)!0.5!(embed)$] {\textbf{Object Condensation} \\ \scriptsize $L = L_{\mathrm{attr}} + L_{\mathrm{rep}} + L_{\beta}$};
    \draw [-Stealth, line width=1.5pt, draw=gray!80] (input) -- (trans);
    \draw [-Stealth, line width=1.5pt, draw=gray!80] (trans) -- (embed);
    \draw [-Stealth, line width=1.5pt, draw=gray!80] (embed) -- (dbscan);
    \draw [-Stealth, line width=1.5pt, draw=gray!80] (dbscan) -- (output);
    \draw [dashed, -Stealth, draw=red!70, line width=1pt] (embed.south) |- (loss.east);
    \draw [dashed, -Stealth, draw=red!70, line width=1pt] (loss.west) -| (trans.south);
    \end{tikzpicture}%
    }
    \caption{The track-reconstruction pipeline}
    \label{fig:pipeline}
\end{figure}

The model is trained on $15\,000$ simulated $\ttbar$ events at $\avgmu = 200$ with a
$60$/$40$ training--validation split, for $100$ epochs, taking approximately three days
on a single NVIDIA A100 GPU. All loss components converge smoothly to a stable plateau.
The full pipeline is shown in Figure~\ref{fig:pipeline}.

At inference the learned embeddings are clustered into track candidates with the
density-based algorithm DBSCAN, which requires no prior specification of the number of
clusters and so adapts to the track multiplicity of each event. Its two hyperparameters,
the neighbourhood radius $\varepsilon$ and the
minimum cluster size $N_{\mathrm{min}}$, are optimized for track reconstruction
efficiency, giving $\varepsilon = 0.05$ and $N_{\mathrm{min}} = 2$. Prior to clustering, a
threshold on the condensation likelihood removes low-confidence triplets as likely fakes,
reducing the combinatorial background.

Each candidate is characterized by the number of unique OT layers it spans, $N_L$.
Requiring $N_L \geq 5$ reproduces the five-layer span of an LST quintuplet and so places
the two algorithms on the same footing; relaxing it to $N_L \geq 4$ probes the
efficiency-versus-purity tradeoff and the recovery of shorter trajectories.

\section{Results}
\label{sec:results}

\subsection{Evaluation procedure}
\label{sec:metrics}

Performance is evaluated on simulated $\ttbar$ events at $\avgmu = 200$ and $\sqrt{s} =
14$~TeV in the Phase-2 detector geometry. Reconstructed candidates are associated to
simulated particles through their constituent mini-doublets: a candidate is matched if
more than $75\%$ of its mini-doublets originate from a single simulated particle, and is
otherwise counted as a misidentified (fake) track. Note that this differs from the
standard CMS convention of using hits to measure purity. Simulated particles from the
hard-scattering vertex with $\pt > 0.8$~GeV, $|\eta| < 2.4$, production-vertex radial
distance $r_{\mathrm{vertex}} < 2.5$~cm and $|z_{\mathrm{vertex}}| < 30$~cm define the
efficiency denominator. Three metrics are reported as functions of $\eta$ and $\pt$: the
\emph{tracking efficiency}, the fraction of simulated particles associated to at least
one reconstructed track; the \emph{fake rate}, the fraction of misidentified
reconstructed tracks; and the \emph{duplicate rate}, the fraction of reconstructed tracks
associated multiple times to the same simulated particle. The efficiency is binned in the
$\eta$ and $\pt$ of the simulated particle, whereas the fake and duplicate rates are
binned in those of the reconstructed candidate; since no track fit is performed, the $\pt$
of a learned candidate is taken to be the $\pt$ estimate of its innermost constituent
triplet. Throughout, the learned candidates ("Transformer OC") are compared with LST
quintuplets as built, before the LST track-candidate step.

\subsection{Condensation likelihood and layer multiplicity}
\label{sec:beta}

\begin{figure}[!t]
    \centering
    \includegraphics[width=0.32\textwidth]{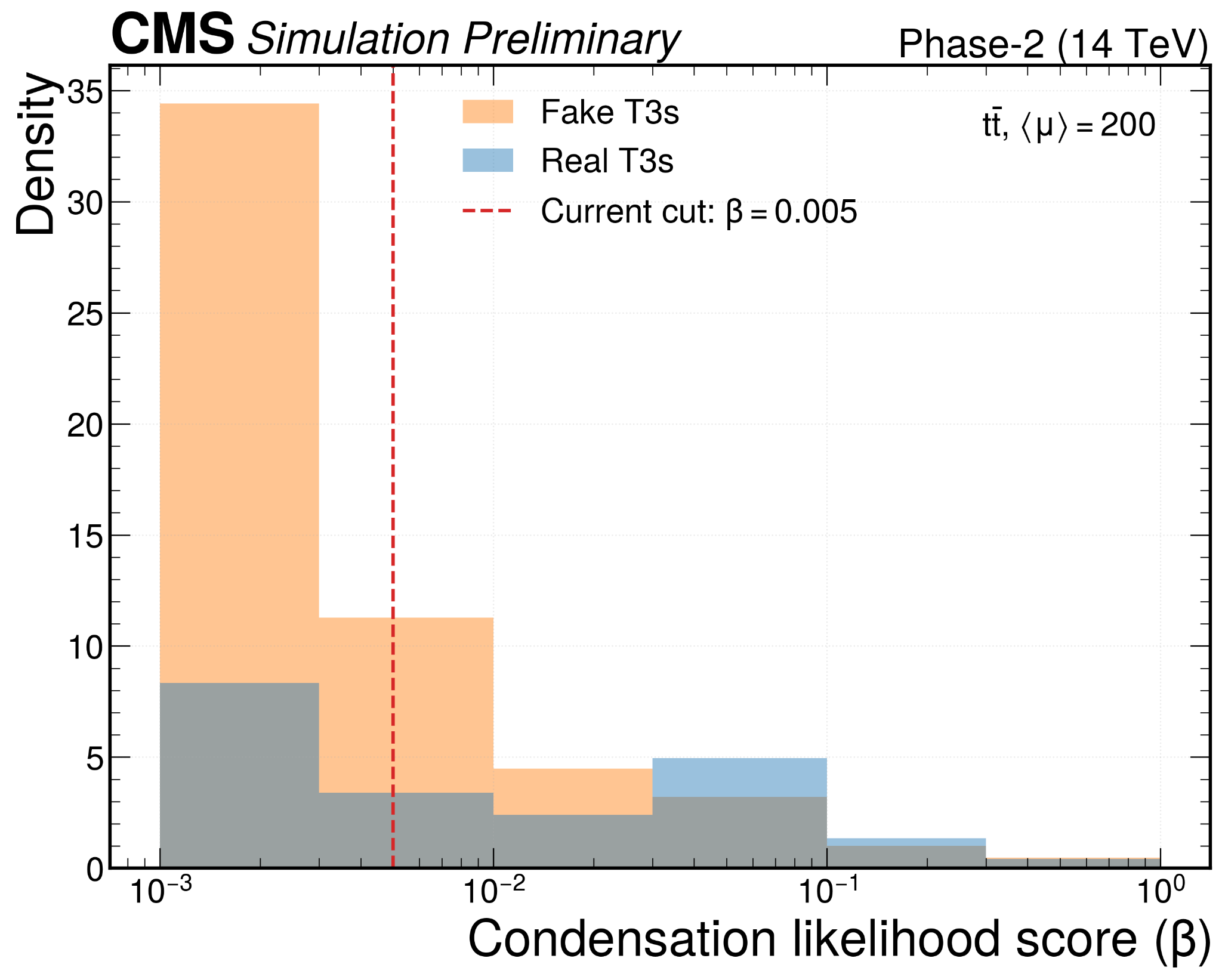}\hfill
    \includegraphics[width=0.32\textwidth]{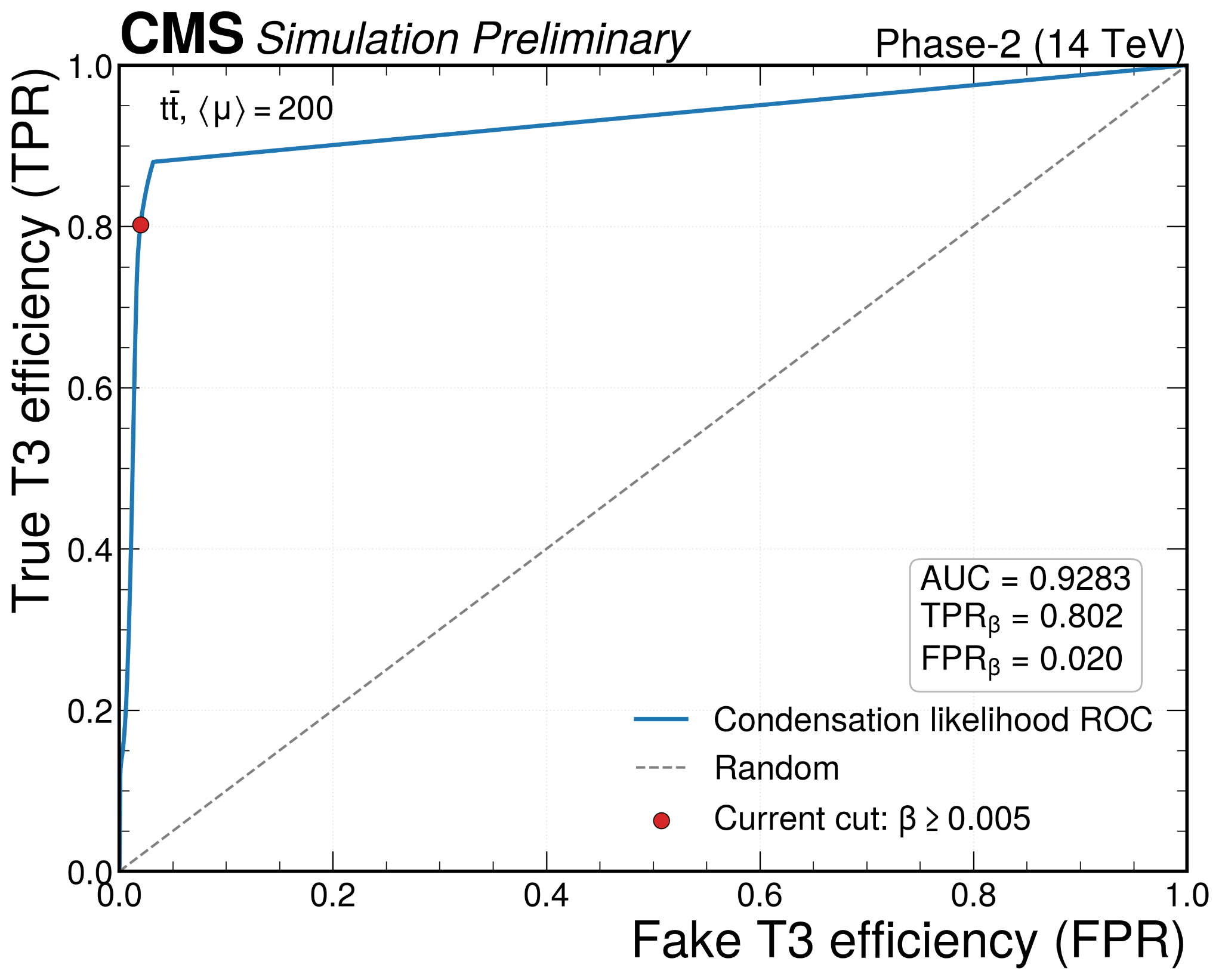}\hfill
    \includegraphics[width=0.26\textwidth]{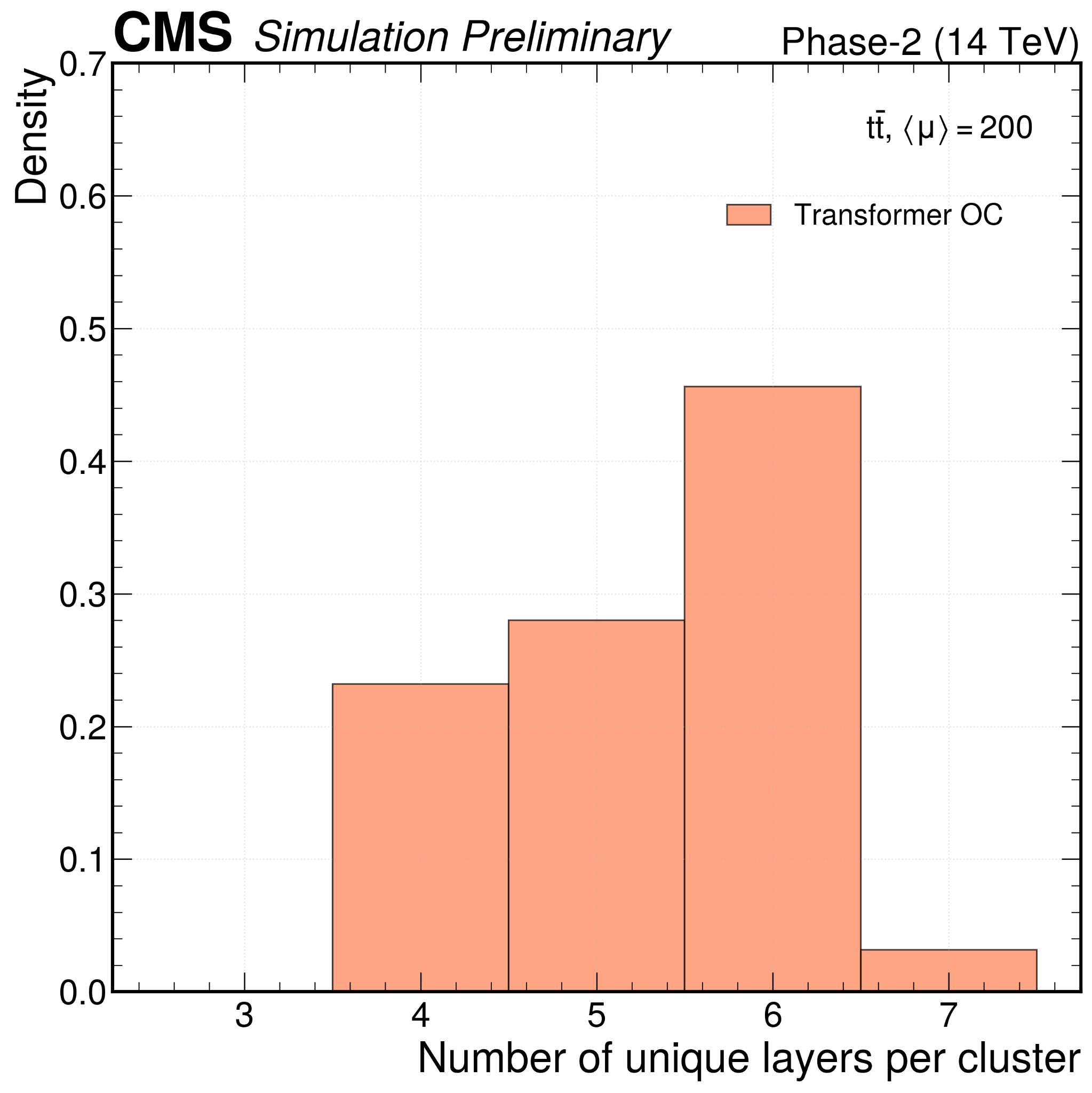}
    \caption{For $\ttbar$ events at $\avgmu = 200$. Left: condensation likelihood $\beta$
    for real and fake T3s, with the operating point $\beta \geq 0.005$ indicated. Centre:
    ROC curve for $\beta$ as a real/fake T3 classifier; the marker is the operating point,
    retaining $80.2\%$ of true triplets at $2.0\%$ fake-triplet efficiency. Right: number
    of unique OT layers $N_L$ spanned by the reconstructed candidates.}
    \label{fig:condensation}
\end{figure}

The condensation likelihood learned by the network separates real from fake triplets
cleanly: fake T3s populate the very lowest scores, so a low threshold suffices. The
chosen operating point, $\beta \geq 0.005$, is set to maximize the efficiency of the
reconstructed clusters under the relaxed criterion $N_L \geq 4$,
rather than the triplet-level classification performance. It
retains roughly $80\%$ of true T3s while accepting only about $2\%$ of fakes,
corresponding to an area under the ROC curve of $0.93$
(Figure~\ref{fig:condensation}, left and centre).

Because the model is free to group any number of triplets into a candidate, the layer
multiplicity of its output is not fixed. Figure~\ref{fig:condensation} (right) shows that
the resulting $N_L$ distribution peaks at six, the full depth of the OT, rather than at
the five layers that a quintuplet spans by construction. The learned clustering therefore
acts as an alternative to the T3-merging step of LST: instead of pairing two triplets
through a shared middle mini-doublet, it groups an unconstrained number of them and so
typically assembles the complete six-layer trajectory.

\subsection{Matched comparison at $N_L \geq 5$}
\label{sec:5layer}

\begin{figure}[!ht]
    \centering
    \includegraphics[width=0.32\textwidth]{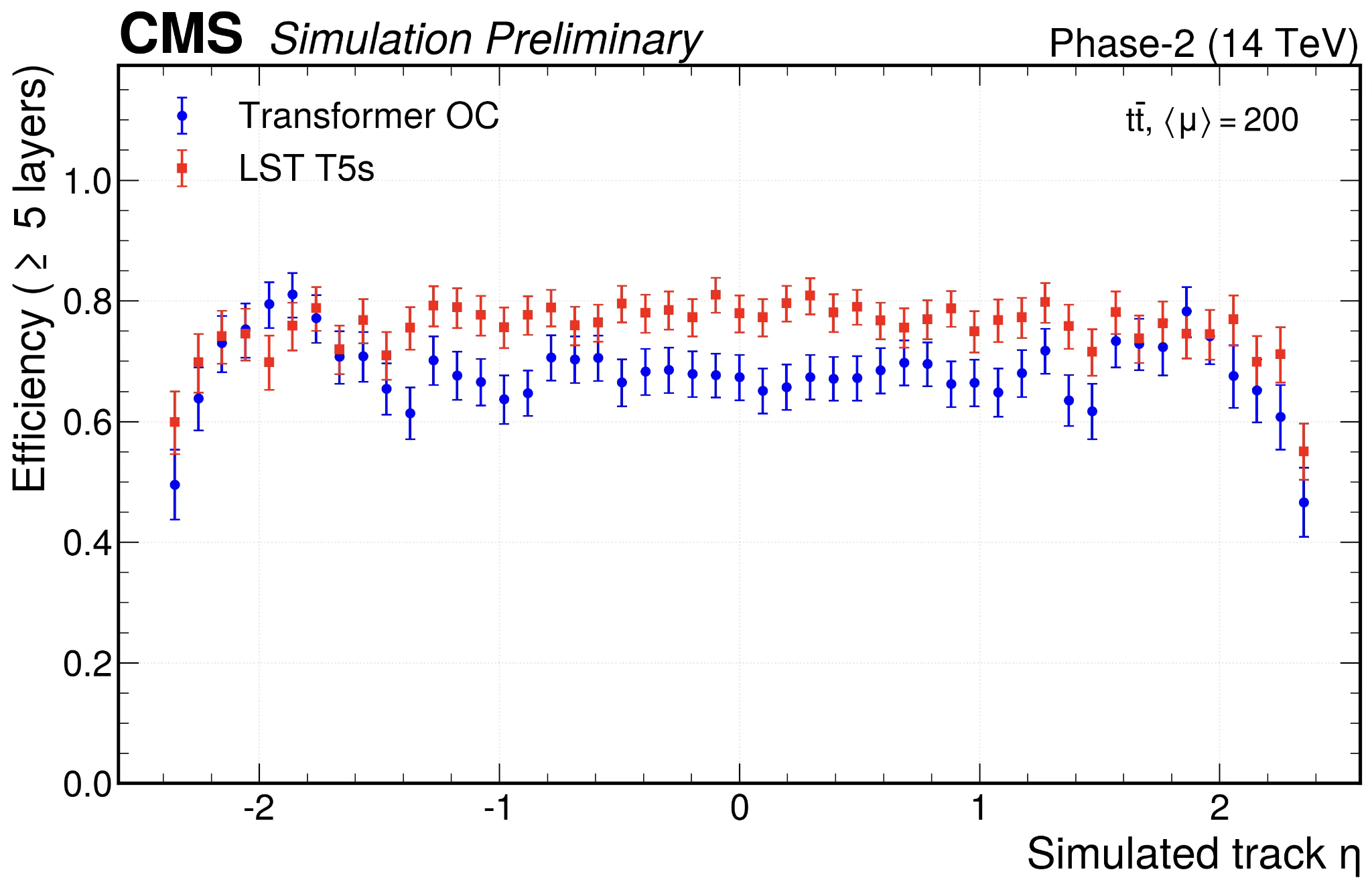}\hfill
    \includegraphics[width=0.32\textwidth]{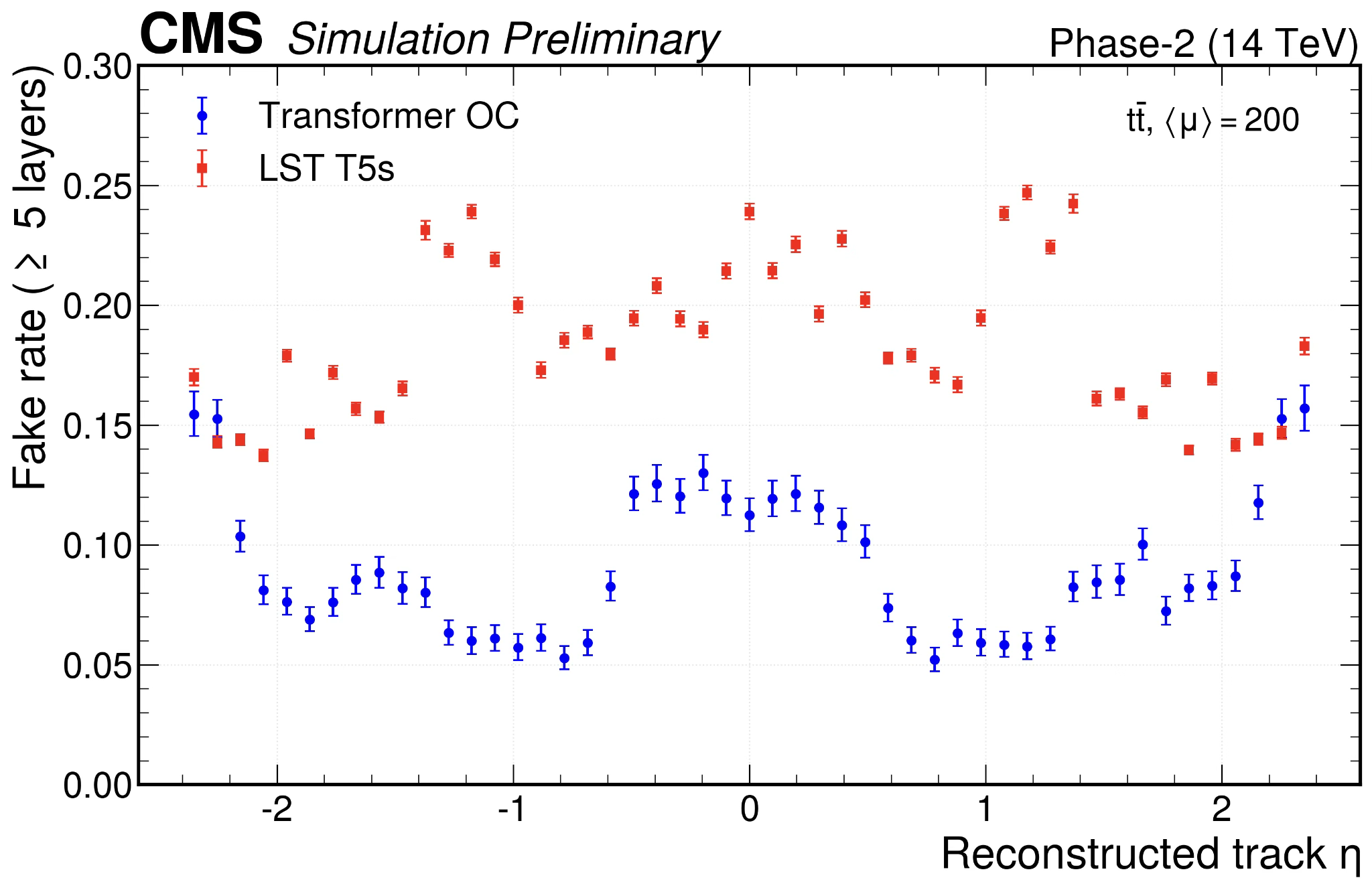}\hfill
    \includegraphics[width=0.32\textwidth]{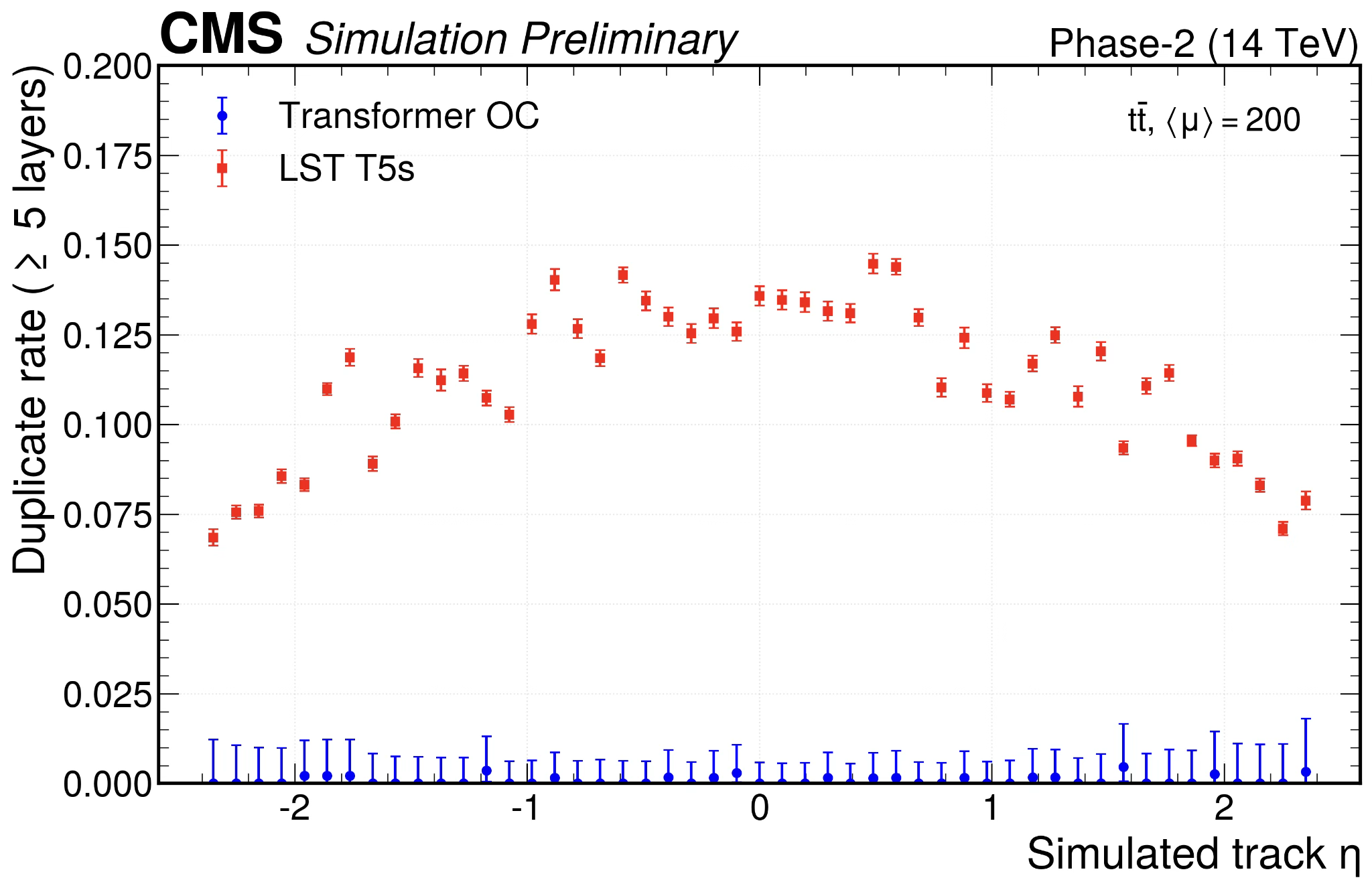}\\[2pt]
    \includegraphics[width=0.32\textwidth]{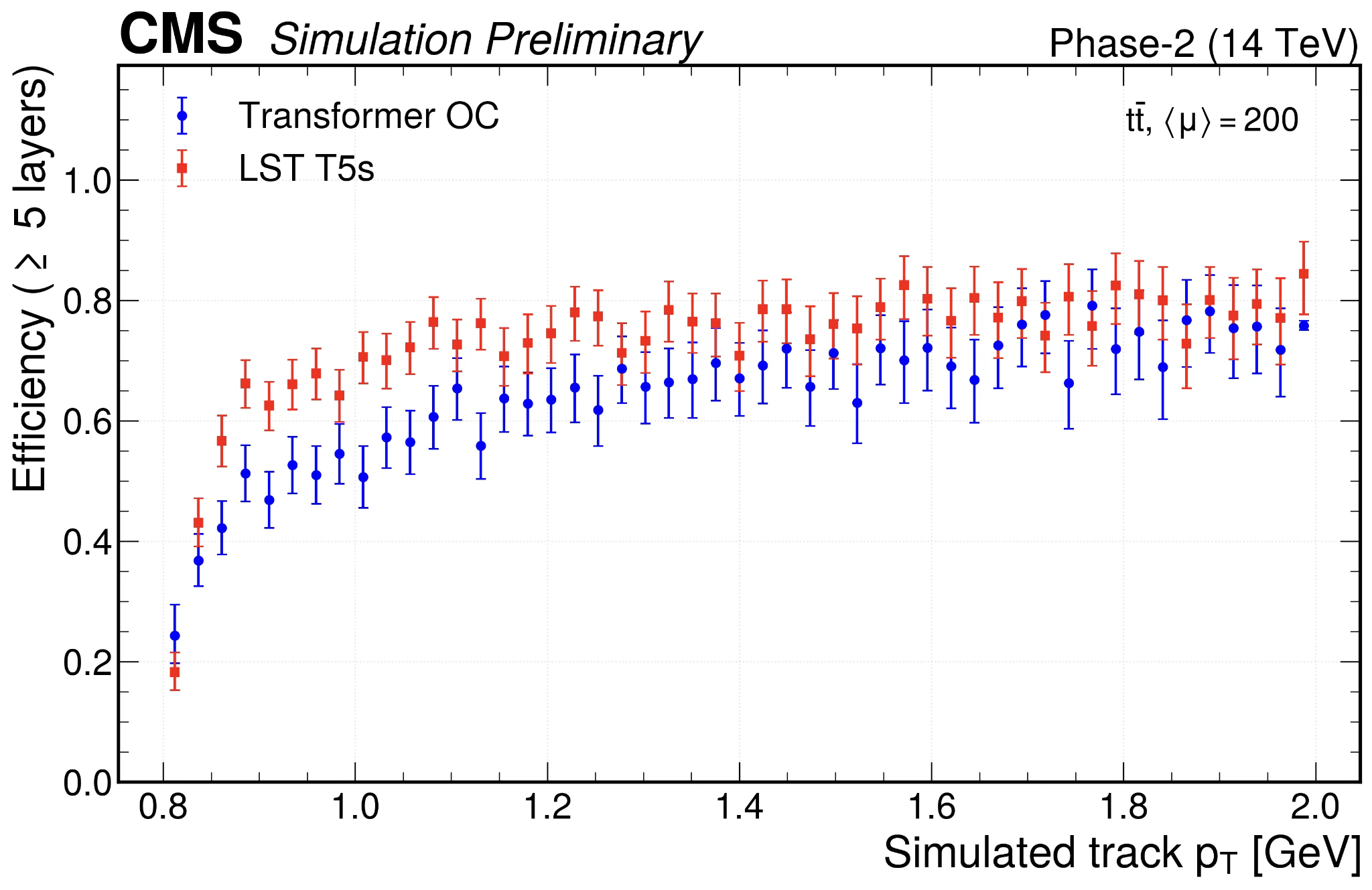}\hfill
    \includegraphics[width=0.32\textwidth]{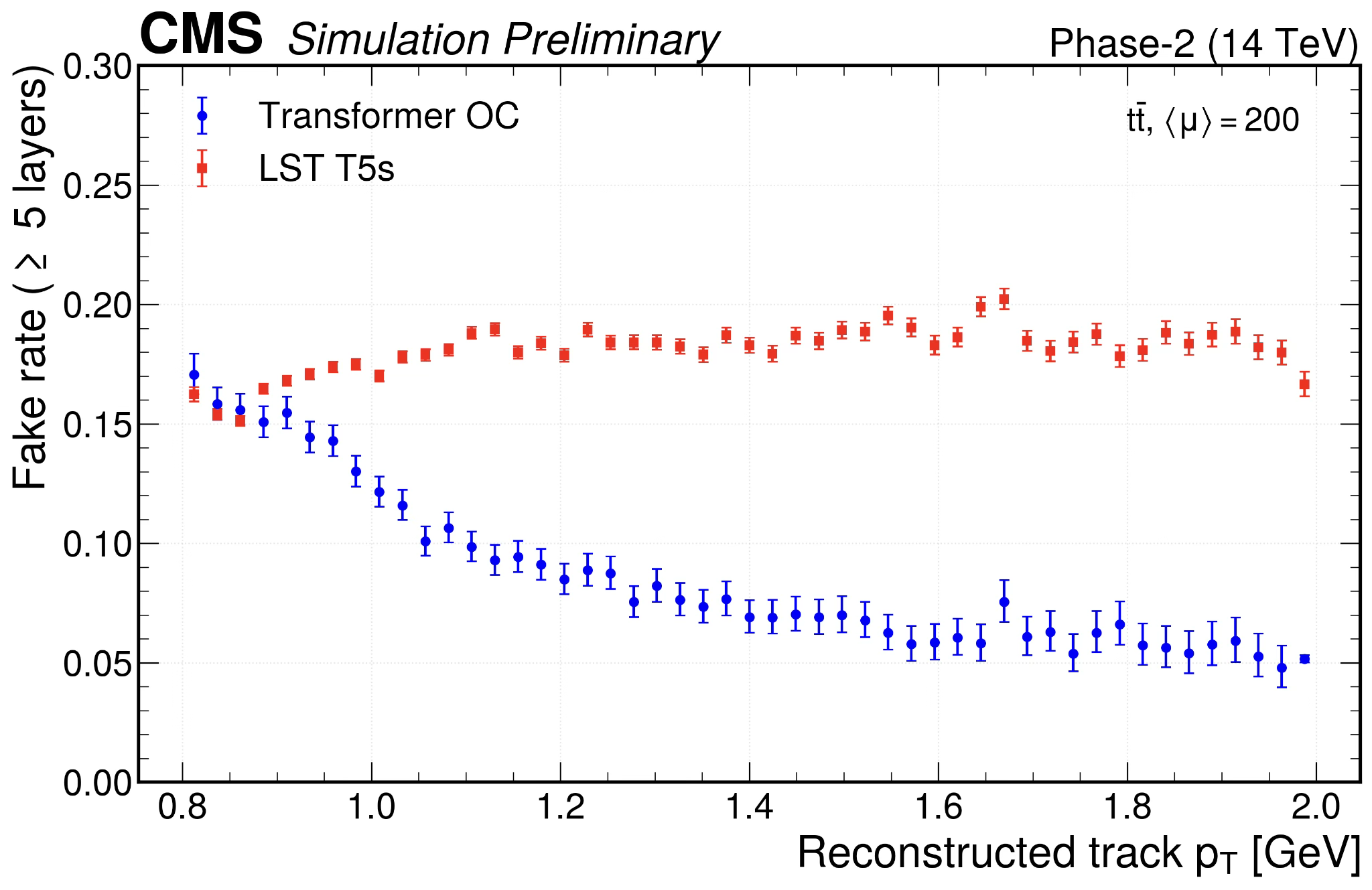}\hfill
    \includegraphics[width=0.32\textwidth]{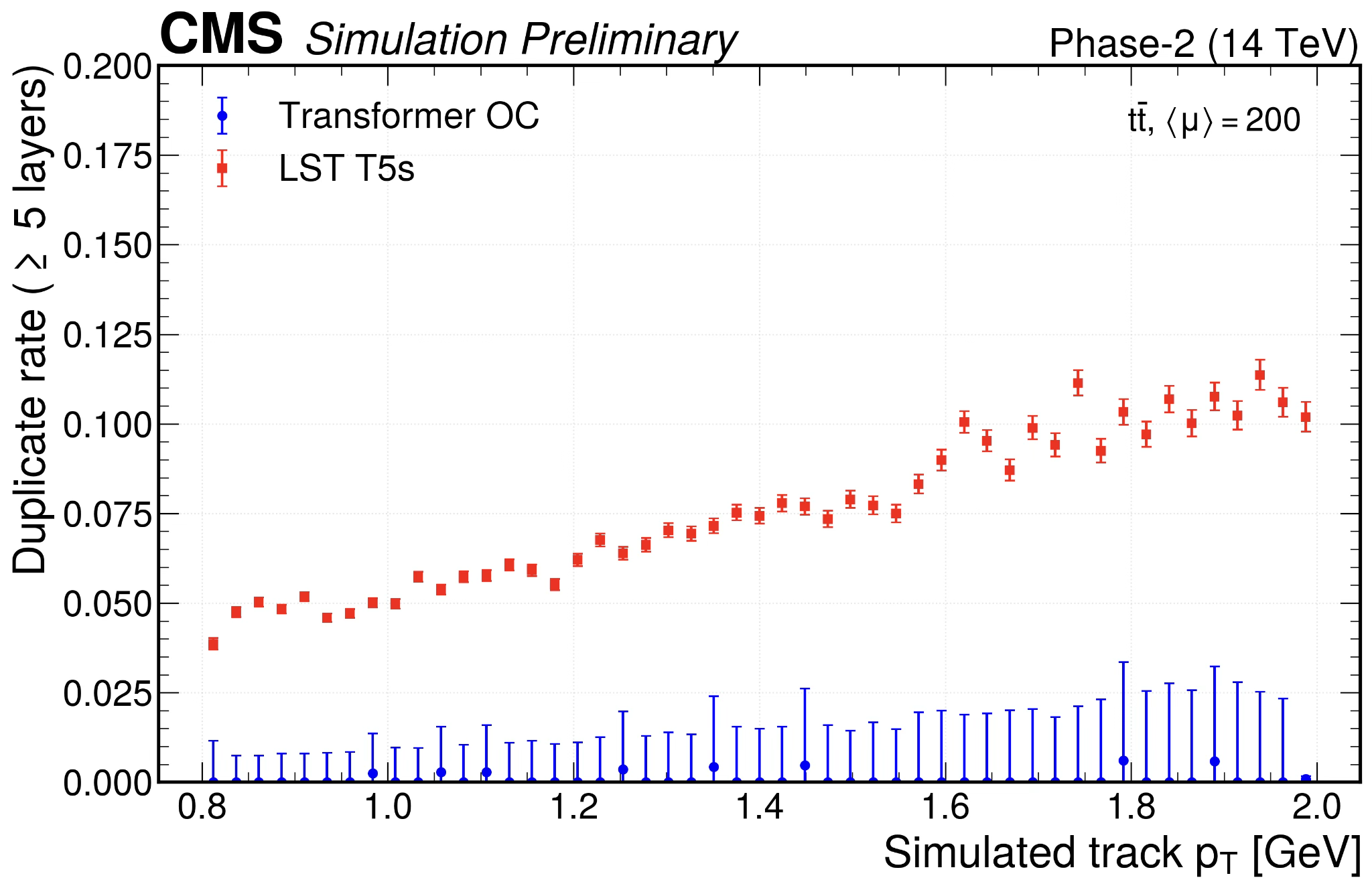}
    \caption{Tracking efficiency (left), fake rate (centre) and duplicate rate (right)
    versus $\eta$ (top) and $\pt$ (bottom) for the Transformer $+$ object-condensation
    candidates (blue) and the LST T5 baseline (red), under the matched $N_L \geq 5$ requirement.}
    \label{fig:perf5}
\end{figure}

At the matched working point the model trades a modest amount of efficiency for a large
gain in purity. The efficiency (Figure~\ref{fig:perf5}, left) sits systematically below the T5
baseline over the core of the acceptance, by roughly ten percentage points, about
$0.67$ against $0.78$ for $|\eta| \lesssim 1.5$, with the gap closing towards $|\eta|
\approx 2$; in $\pt$ the deficit is largest at the low end and the two converge to within
their uncertainties above roughly $1.4$~GeV. The purity, on the other hand, improves
markedly. The fake rate falls steadily with $\pt$, from a value comparable to the baseline
near $0.8$~GeV to about $0.05$ at $2$~GeV, while the T5 fake rate stays flat at
$0.18$--$0.19$; across $\eta$ the model sits at $0.05$--$0.13$ against $0.15$--$0.26$, so
it is roughly three times cleaner over most of the range.

The duplicate rate is the most striking difference: it is consistent with zero
everywhere, whereas the T5 duplicate rate climbs from $0.04$ to above $0.10$ with
increasing $\pt$ and reaches $0.15$ in the central barrel. The difference arises from the fact that
LST enumerates every valid triplet pairing, so one particle can yield several overlapping
quintuplets, removed only by a later dedicated step~\cite{CMS-DP-2025-048} that is not
applied here. Clustering instead partitions the triplets, and the object-condensation loss
drives all triplets of a particle onto a single condensation point, so a particle can give
rise to only one candidate.

\subsection{Relaxed requirement at $N_L \geq 4$}
\label{sec:4layer}

\begin{figure}[!ht]
    \centering
    \includegraphics[width=0.32\textwidth]{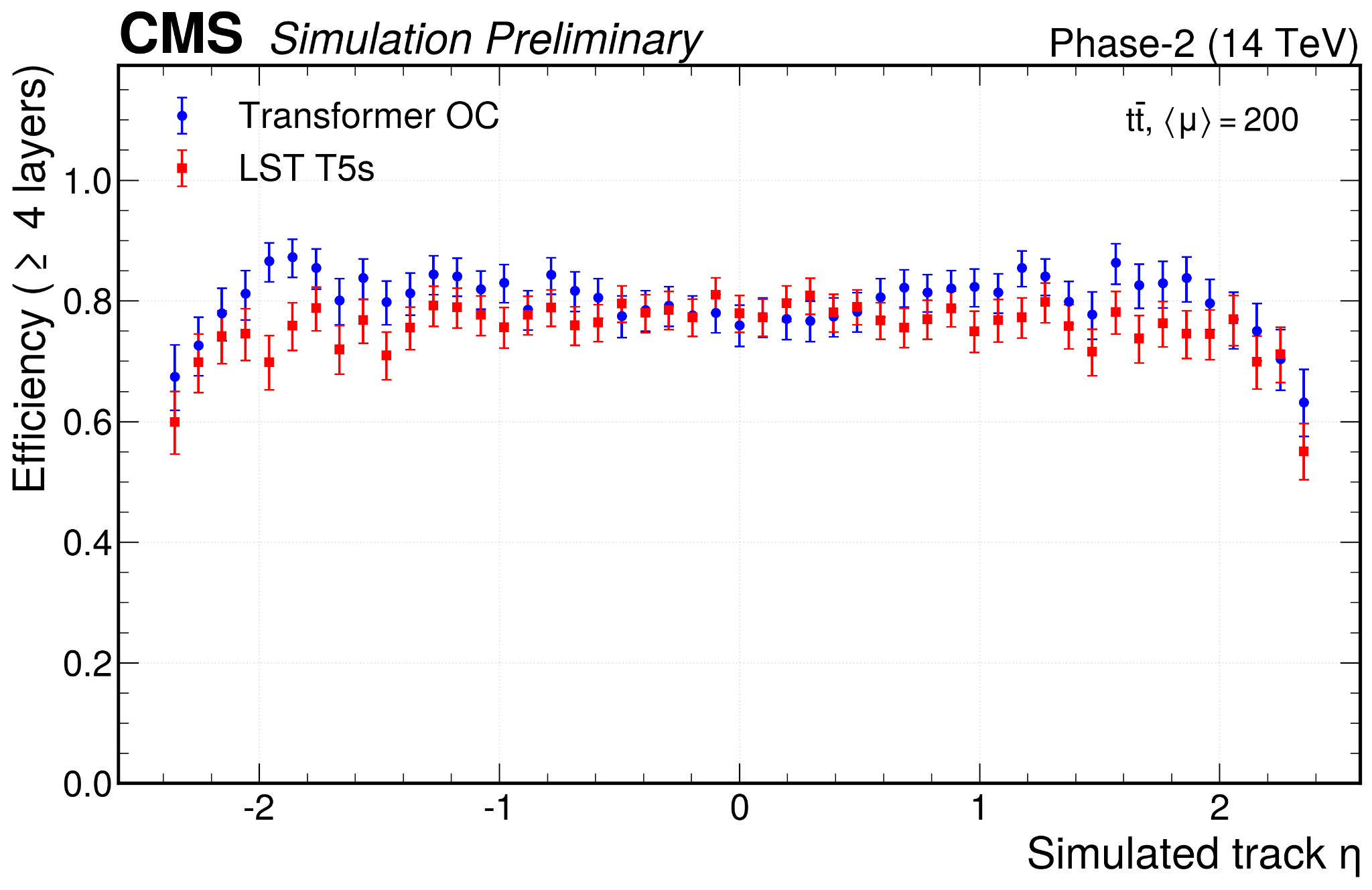}\hfill
    \includegraphics[width=0.32\textwidth]{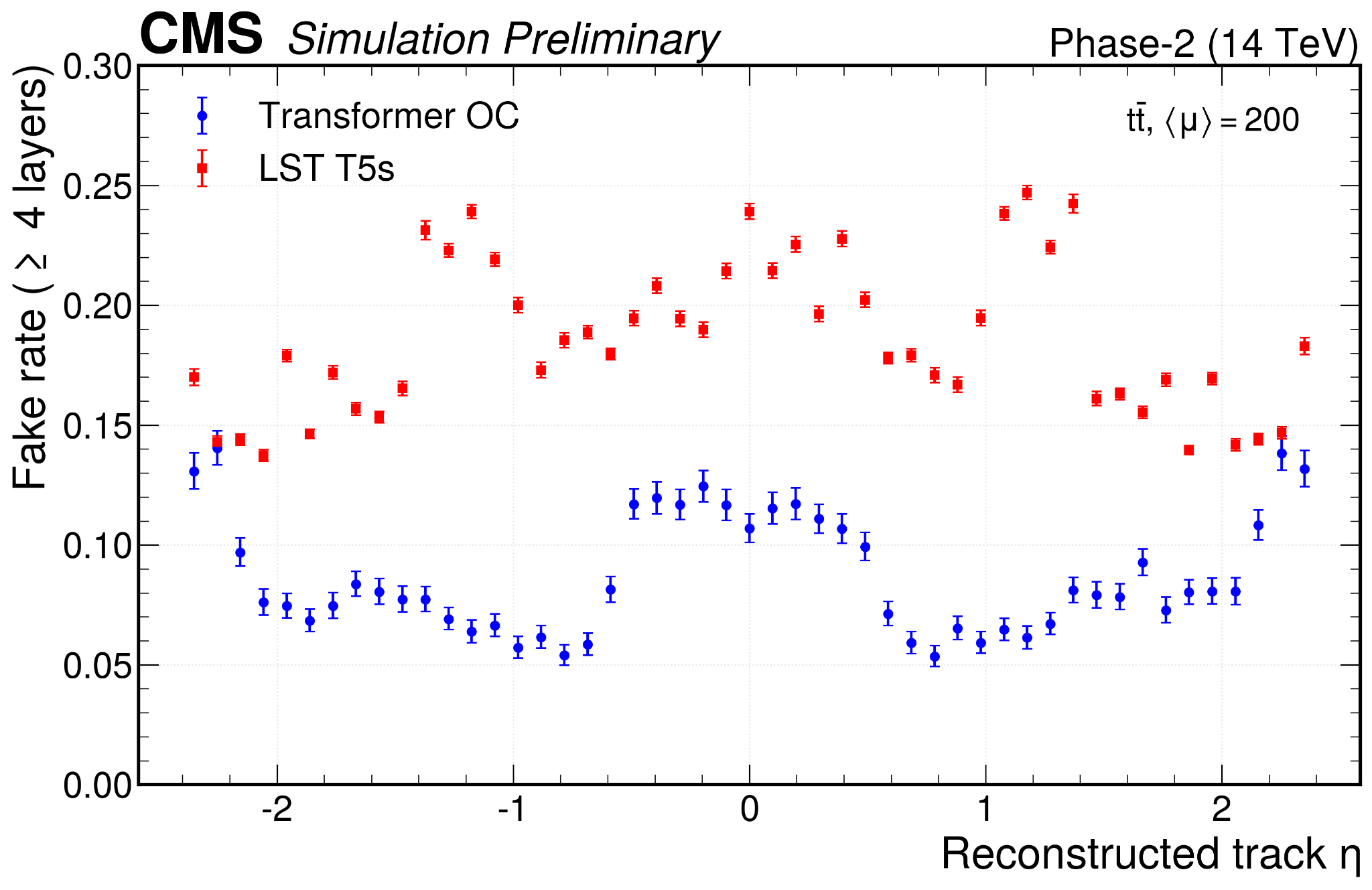}\hfill
    \includegraphics[width=0.32\textwidth]{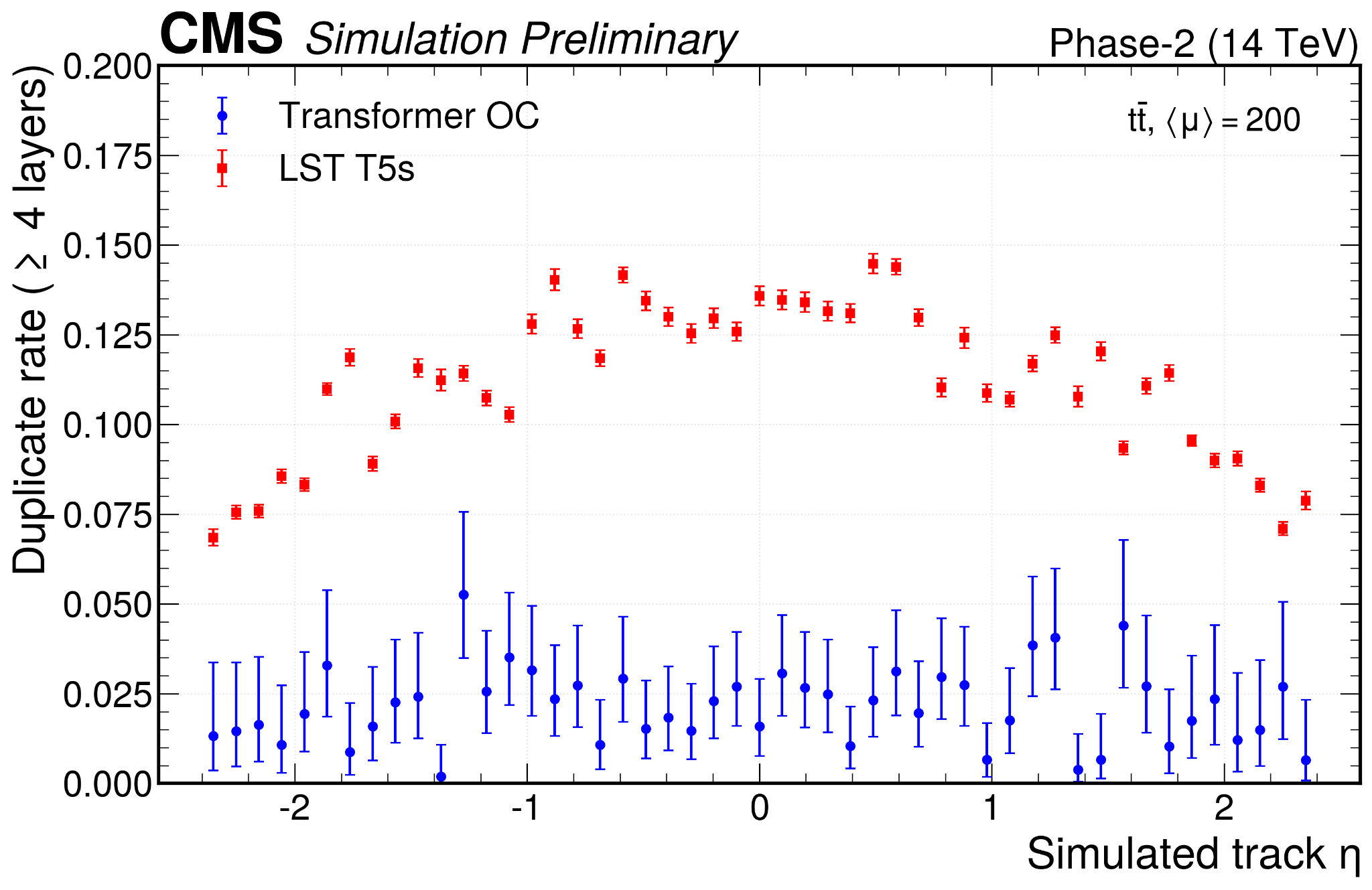}\\[2pt]
    \includegraphics[width=0.32\textwidth]{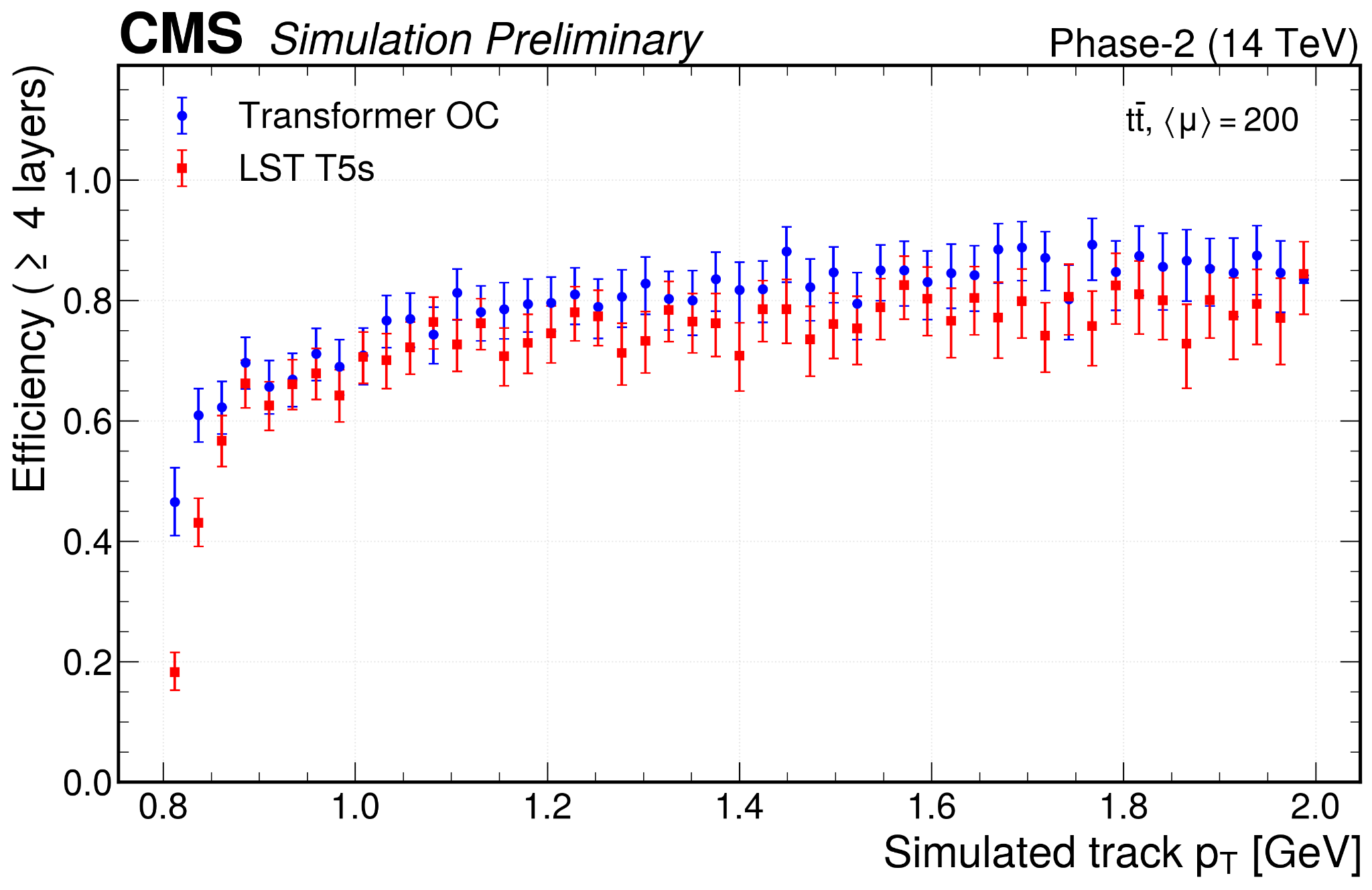}\hfill
    \includegraphics[width=0.32\textwidth]{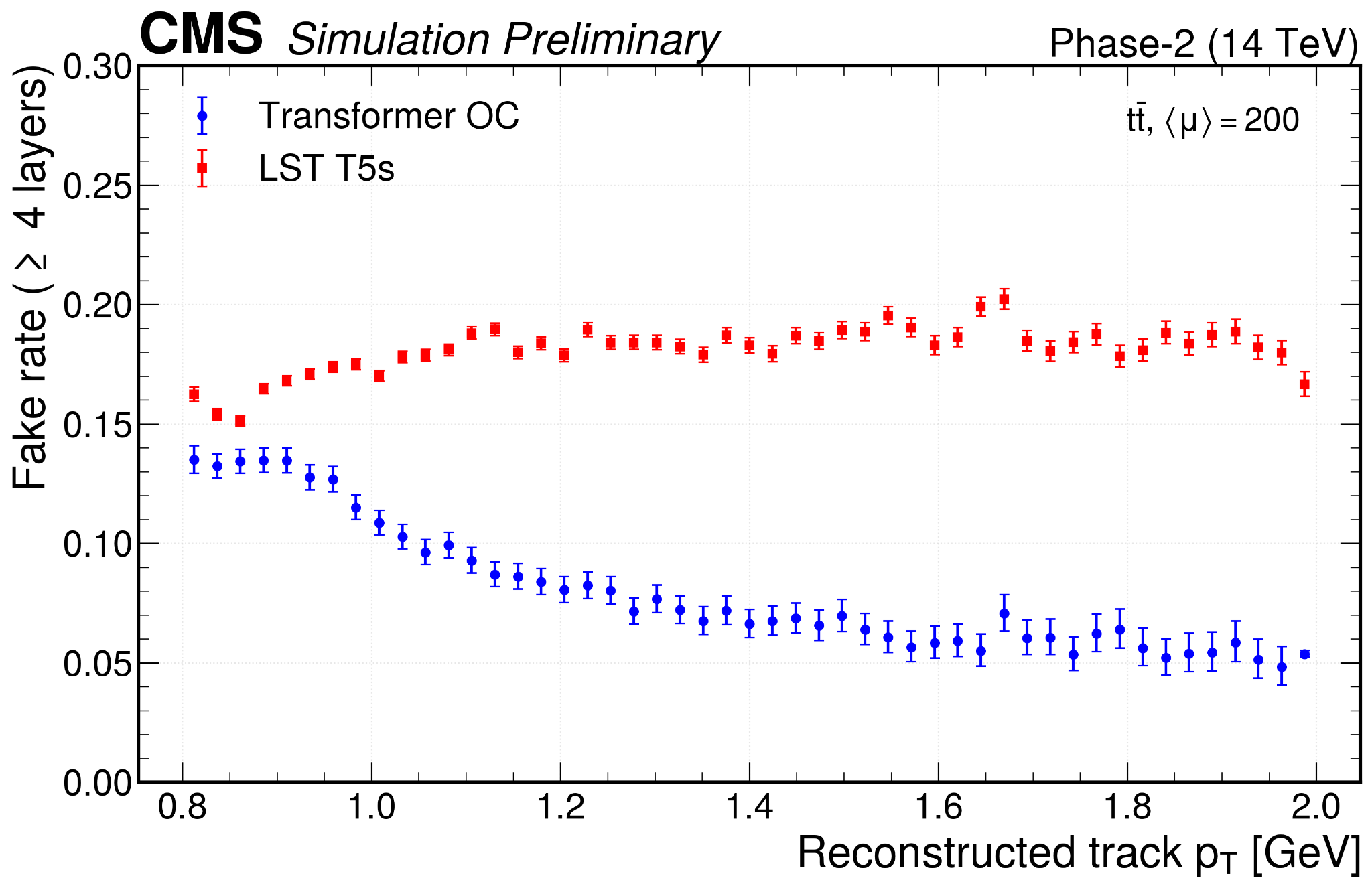}\hfill
    \includegraphics[width=0.32\textwidth]{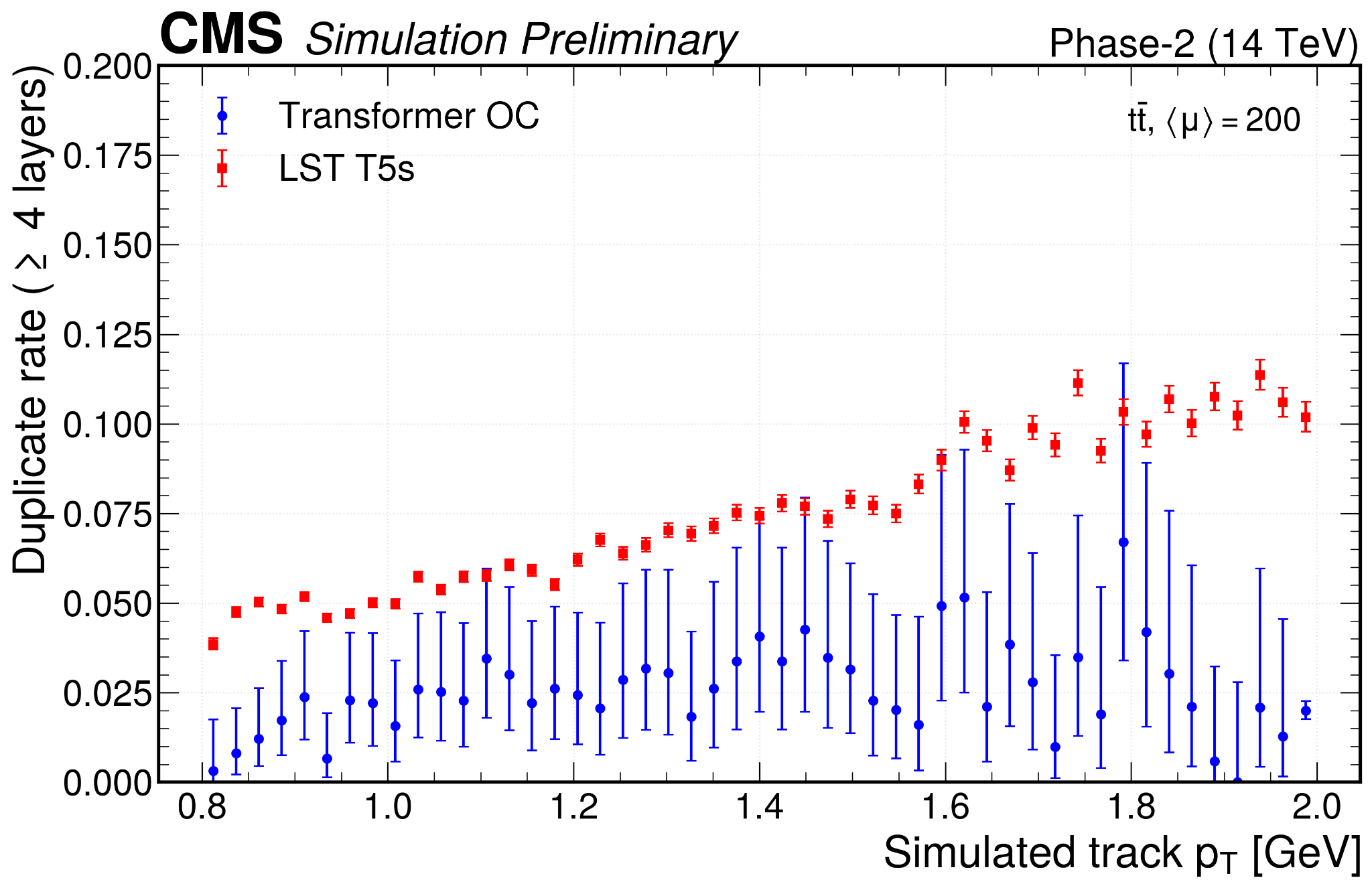}
    \caption{As Figure~\ref{fig:perf5}, but with the relaxed $N_L \geq 4$ requirement
    applied to the Transformer $+$ object-condensation candidates; the LST T5 baseline is
    unchanged.}
    \label{fig:perf4}
\end{figure}

Relaxing the requirement to $N_L \geq 4$ recovers the shorter trajectories that the matched
selection discards, and with them the efficiency (Figure~\ref{fig:perf4}). The model then
matches or exceeds the T5 baseline across the full range, reaching $0.80$--$0.87$ against
$0.70$--$0.80$, with the gain concentrated at $1 \lesssim |\eta| \lesssim 2$. The fake rate
is essentially unchanged with respect to the five-layer selection, and the duplicate rate,
while no longer identically zero, stays at $0.01$--$0.04$ against $0.05$--$0.14$.

This is arguably the more representative operating point for a learned algorithm whose
candidates are not constrained to a fixed layer span, and it indicates that the deficit
seen at $N_L \geq 5$ reflects the layer-count requirement rather than a failure of the
embedding.

\section{Summary and outlook}
\label{sec:summary}

We have presented the first CMS demonstration of a machine-learning track-reconstruction
model built on Transformers and object condensation~\cite{CMS-DP-2026-030}. Rather than
learning from raw hits, the model takes LST triplets as tokens, so that the geometry and
local momentum already encoded in each triplet enter the network as a strong physics
prior. A geometrically masked, sparse self-attention encoder embeds the triplets under an
object-condensation objective, and DBSCAN forms the final track candidates in a single
inference step, without any prior on their number.

Evaluated on $\ttbar$ events at $\avgmu = 200$, the model achieves a fake rate roughly
three times lower than the LST quintuplet baseline and a duplicate rate consistent with
zero, and when shorter trajectories are admitted ($N_L \geq 4$) it matches or exceeds the
baseline efficiency across the full acceptance. These results establish LST objects as an
effective, physics-rich tokenization for learned tracking, and show that a
physics-informed model can match, and in places surpass, a state-of-the-art combinatorial
algorithm under the harshest conditions foreseen at the HL-LHC.

This work is a standalone algorithmic demonstration and should be viewed independently of
official CMS tracking benchmarks. The natural next steps are to integrate the model into
the CMS reconstruction and trigger software alongside LST, to measure its throughput
against the latency budget of the Phase-2 HLT, and to extend it from the Outer Tracker towards the
full tracking volume.

{\scriptsize
\bibliography{refs}
}

\end{document}